\documentclass[prd,preprint,showpacs,showkeys,nofootinbib,superscriptaddress]{revtex4}
\usepackage{feynmf}
\usepackage{graphicx}  %
\usepackage{amsmath}

\newcommand{\simle}{\hspace*{0.2em}\raisebox{0.5ex}{$<$}
\hspace{-0.8em}\raisebox{-0.3em}{$\sim$}\hspace*{0.2em}}

\newcommand{\bqa}{\begin{eqnarray}} \newcommand{\eqa}{\end{eqnarray}}
 
\def\mqo2{{\!\!\!}}

\newcommand{\bce}{\begin{center}} \newcommand{\ece}{\end{center}}
\newcommand{\beq}{\begin{equation}} \newcommand{\eeq}{\end{equation}}
\newcommand{\bea}{\begin{eqnarray}}\newcommand{\eea}{\end{eqnarray}}
\newcommand{\beqy}{\begin{eqnarray}}\newcommand{\eeqy}{\end{eqnarray}}
\newcommand{\ub}{\bar{u}} \newcommand{\pp}{p^{\prime}}
 
\newcommand{\gn}{\gamma^{\nu}} \newcommand{\gm}{\gamma^{\mu}}
\newcommand{\ga}{\gamma^{\alpha}} \newcommand{\gda}{\gamma_{\alpha}}
\newcommand{\gf}{\gamma_5} \newcommand{\gb}{\gamma^{\beta}}
\newcommand{\sla}{{\hskip -1.75mm}\slash}    \newcommand{\Tr}{{\rm Tr}} \def\ni{\noindent} \input epsf
\begin{document}

\title{Muon Production in Low-Energy\\
Electron-Nucleon and Electron-Nucleus Scattering}

\author{Prashanth Jaikumar}
\affiliation{Department of Physics and Astronomy, Ohio University, 
Athens OH 45701 USA}
\affiliation{The Institute of Mathematical Sciences, C.I.T Campus, Taramani, 
Chennai 600113, India}
\author{Daniel R. Phillips}
\affiliation{Department of Physics and Astronomy, Ohio University, 
Athens OH 45701 USA}
\author{Lucas \surname{Platter}}
\affiliation{Department of Physics and Astronomy, Ohio University, 
Athens OH 45701 USA}
\author{Madappa Prakash}
\affiliation{Department of Physics and Astronomy, Ohio University, 
Athens OH 45701 USA}

\begin{abstract}
Recently, muon production in electron-proton scattering has been
suggested as a possible candidate reaction for the identification of
lepton-flavor violation due to physics beyond the Standard Model. Here
we point out that the Standard-Model processes $e^- p \rightarrow
\mu^- p \bar{\nu}_\mu \nu_e$ and $e^- p \rightarrow e^- n \mu^+
\nu_\mu$ can cloud potential
beyond-the-Standard-Model signals in $ep$ collisions. We find that
Standard-Model $e p \rightarrow \mu X$ cross sections
exceed those from lepton-flavor-violating operators by
several orders of magnitude. We also discuss the possibility of using
a nuclear target to enhance the $e p \rightarrow \mu X$ signal.
\end{abstract}

\maketitle

\section{Introduction}
        \label{sec_intro}

A number of experiments over the past decade provide compelling
evidence that the neutrino mass matrix is non-diagonal in the basis of
weak eigenstates $|\nu_{\alpha}\rangle; \alpha=e,\mu,~{\rm and}~\tau$
(see~\cite{Fogli} for a recent review).  This knowledge has led to
renewed interest in lepton-flavor violation (LFV), which can be probed
by searches for rare decays such as $\mu\rightarrow e\gamma$.  Such
LFV decays are possible when the Standard Model is extended to include
neutrino mass and neutrino mixing, but the resulting cross section is
exceedingly small (branching ratio, ${\rm BR} \sim 10^{-60}$) as the
process scales with the fourth power of the ratio of the neutrino mass
to the $W$-boson mass~\cite{Diener:2004kq}. However, a significantly
larger branching ratio, ${\rm BR} \sim 10^{-12}$, results from the
Minimal Super-Symmetric extension of the Standard Model
(MSSM)~\cite{Blazek:2004cg}. The MEG
(\underline{m}u$\rightarrow$\underline{e} \underline{g}amma)
experiment at the Paul Scherer Institute (PSI)~\cite{Ritt:2006cg} is
capable of detecting branching ratios as small as $10^{-13}$ at a 90\%
confidence level, and will search for the LFV decay $\mu\rightarrow
e\gamma$.  Such searches for lepton-flavor violation potentially offer
an intriguing window on beyond-the-Standard-Model (BSM) physics.

Recently, the possibility of observing lepton-flavor violation in
fixed-target electron scattering has been raised as an
alternative~\cite{Diener:2004kq} to experiments searching for the rare
$\mu \rightarrow e \gamma$ decay.  Facilities with electron beams of
high intensity and significant duty factor, such as Jefferson Lab,
seem to be natural places to perform experiments to search for $e p
\rightarrow \mu p$.  Hereafter, we will refer to this
\underline{e}lectron-to-\underline{mu}on conversion process as EMU.
However, the conclusion of Ref.~\cite{Diener:2004kq} is that even
under the most favorable dynamical scenario (a heavy right-handed
Majorana neutrino with $m_{\nu}\sim{\cal O}(m_W))$, the Standard-Model
supplemented by dynamics that results in neutrino oscillations yields
a cross section $\sigma\approx 10^{-27}$ femtobarns (fb) for EMU---so
low as to be inaccessible to current experiments.

In this paper, we discuss two Standard-Model processes that can cloud
an EMU signal in $e^- p$ scattering by generating final states $\mu^-
X$ other than the desired final state $\mu^- p$.  In Standard-Model
mechanisms, the additional particles $X$ must have baryon number 1,
muon lepton number --1, and electron lepton number +1.  Two such
reactions are: (i) $e^- p \rightarrow \mu^- \bar{\nu}_\mu \nu_e p$,
and (ii) $e^-p \rightarrow e^-n\mu^+\nu_\mu$. From now on, we refer to
these reactions as ``muon-production processes'' in order to
distinguish them from EMU.

The first reaction involves only electroweak interactions, and takes
place as the electron goes off-shell in the scattering event by an
amount corresponding to the momentum of the virtual photon exchanged
with the target.  The electron then decays via the weak interaction to
a muon, accompanied by the emission of $\nu_e$ and
$\overline{\nu}_{\mu}$.  We note that this process can also take place
off a neutron, although in practice this implies a nuclear target.  In
fact, for the nuclear-target case coherent electron interactions with the total
nuclear charge can enhance the signal.

The second muon-production process involves the strong
interaction. Even when the electron energy is below the
pion-production threshold the exchanged virtual photon can interact with
the ``cloud'' of virtual pions that surrounds the nucleon. This can
generate an off-shell $\pi^+$, which decays to a $\mu^+$ and
$\nu_\mu$.  (In electron-nucleus scattering, the presence of neutrons
allows $e^-n \rightarrow e^-p\mu^-\bar\nu_\mu$ to occur through
virtual $\pi^-$'s.)  As the energy of the incident electron approaches
the pion threshold, the time for which the virtual pion lives (and
hence the distance it travels before decaying)
increases. Consequently, these muon-production reactions switch over
to pion electro-production at electron energies of about 140 MeV
swamping any possible EMU signals.

This paper is organized as follows. Section~\ref{sec-BSM} presents a
pedagogical calculation of electron-proton scattering as a probe of
BSM physics using generic low-energy effective couplings that can
cause EMU. The diagram involving photon exchange plays a dominant
role, but is severely constrained by the experimental bound on the
coupling obtained from the $\mu \to e \gamma$ process. (For a specific
realization in the MSSM see Ref.~\cite{Blazek:2004cg}.) In
Sec.~\ref{sec-EMcrosssection}, we derive the matrix element for the
process $e^- p \rightarrow \mu^- \bar{\nu}_\mu \nu_e p$ from
electroweak theory, compute the size of the cross section, and provide
a simple explanation for the order-of-magnitude of our result.  In
Sec.~\ref{sec-strongcrosssection}, the virtual-pion production and
decay contributions to the matrix element for $ep \rightarrow en \mu^+
\nu_\mu$, and the resulting differential cross section are presented.
Due to differences in the interaction couplings and phase-space
factors, the cross section for $e^- p \rightarrow e^- n \mu^+ \nu_\mu$
turns out to be several orders of magnitude larger than that for $e^-
p \rightarrow \mu^- \bar{\nu}_\mu \nu_e p$.  Therefore any EMU
experiment seeking BSM (or even electroweak) physics would either have
to veto processes in which a scattered electron is detected in
coincidence with the produced muon, or, more feasibly, detect the
charge of any muons produced in the electron-proton collision.  In
Sec.~\ref{sec-nucleus}, we describe muon-production in
electron-nucleus scattering, including the relative importance of
collective nuclear excitations.  We present our summary and
conclusions in Sec.~\ref{sec-conclusion}. Details of phase-space
integrations and numerics are provided in the appendices.


\section{Electron-muon conversion via physics beyond the standard
  model}

\label{sec-BSM}

In this section, we consider the differential cross section for the
reaction $e p \rightarrow \mu p$ induced by operators which change
lepton flavor, and hence are low-energy manifestations of physics
``beyond the Standard Model'' (BSM). All such
operators are, by definition, dimension five or above, and so they
produce cross sections suppressed by (at least) one power of
$m_\mu/\Lambda$, where $\Lambda$ is the scale of the physics that
results in lepton-flavor violation. We will show that there
is a dimension-five operator that could, in principle, produce a
sizeable $e p \rightarrow \mu p$ cross section. However, in practice, bounds
from the non-observation of the process $\mu \rightarrow e \gamma$
preclude any observable muon production via this dimension-five BSM
operator.
\begin{figure}[t]
\includegraphics[height=5cm]{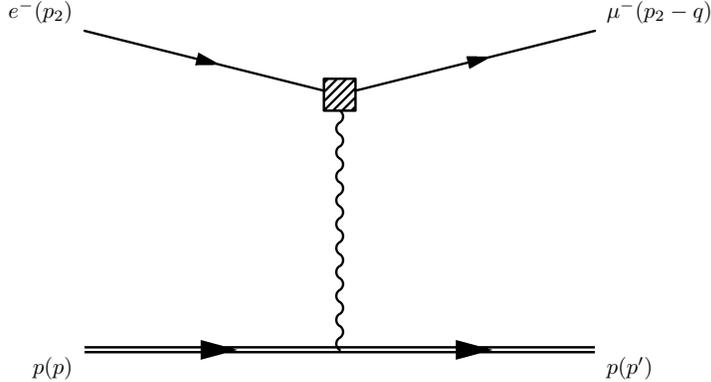}
\caption{\label{susy} Beyond-the-Standard-Model contribution to muon
  production (``electron-muon conversion''). The hatched vertex
  represents the dimension-five coupling of Eq.~(\ref{eq:LBSM1}). The
  solid lines denote leptons, while the double line is the proton. The
  particle momenta are indicated in parentheses.}
\end{figure}

The BSM operator associated with the decay $\mu \rightarrow e
\gamma$ can be written as
\begin{equation}
{\cal L}_I=-\frac{e v}{\Lambda^2} (\bar{\psi}_e \sigma_{\alpha \beta}
F^{\alpha \beta} \psi_\mu + \rm{h.c.})\,,
\label{eq:LBSM1}
\end{equation}
where $\psi_l$ is the lepton field of family $l$ ($e~{\rm or}~\mu$),
and $F^{\alpha \beta}$ is the electromagnetic field-strength
tensor. The object $v$ is the Higgs vacuum expectation value, and
$\Lambda$ is the scale of the BSM physics that induces this operator.
The Higgs vacuum expectation value appears because while the
operator is of dimension five, it is suppressed by an additional power
of $v/\Lambda$ because it changes lepton chirality. Dimension-six BSM
structures which have the low-energy form
\begin{equation}
{\cal L}_I^{\rm contact} \sim \frac{e}{\Lambda^2} \bar{\psi}_e O \psi_\mu
\bar{N} O N \,,
\label{eq:LBSM2}
\end{equation}
where $N$ denotes the nucleon field and the $O$'s are operators
(potentially with Lorentz indices that are contracted with one
another) can also appear in ${\cal L}_I$. However, their effects are
suppressed relative to the operator in Eq.~(\ref{eq:LBSM1}).

The Feynman diagram for $e p \rightarrow \mu p$ for the coupling in
Eq.~(\ref{eq:LBSM1}) is shown in Fig.~\ref{susy}. The general
form of the nucleon current (given parity invariance,
time-reversal invariance and gauge invariance) can be parameterized
using two functions $F_1$ and $F_2$:
\begin{equation}
\langle j^\mu \rangle=e\bar{u}_N({\bf p}')\left [F_1(Q^2) \gamma^\mu + 
\kappa F_2(Q^2)
\frac{i}{2M} \sigma^{\mu \nu} q_\nu\right] u_N({\bf p}) \,,
\label{eq:nuccurrent}
\end{equation}
where $F_1(Q^2)$ and $F_2(Q^2)$ are the Dirac and Pauli nucleon form
factors respectively, $\kappa$ is the proton's anomalous magnetic
moment, $M$ is the proton mass, and $q=p'-p$, with $Q^2=-q^2 > 0$.
Employing this to evaluate the matrix element associated with
the diagram in Fig.~\ref{susy}, we find
\begin{equation}
{\cal M}=\frac{e^2 v}{\Lambda^2} \frac{1}{-q^2} \bar{u}_\mu({\bf p}_2')
[\gamma_\alpha,\not \! q] u_e({\bf p}_2) \bar{u}_N({\bf p}') \left[F_1(Q^2)
\gamma^\alpha + \kappa \frac{F_2(Q^2)}{4M} [\gamma^{\alpha},\not \! q]\right]
u_N({\bf p})\,,
\end{equation}
where now $q=p_2'-p_2=p-p'$ is the four-momentum of the virtual photon that is
exchanged.

The spin-summed-and-averaged squared matrix element can be written as
\begin{equation}
\overline{|{\cal M}|^2}=\frac{e^4 v^2}{\Lambda^4} L^{\alpha \beta} H_{\alpha
\beta} \frac{1}{(q^2)^2}\,,
\end{equation}
where the lepton and hadron tensors are both transverse with respect to the
photon four-vector $q$, that is, 
\begin{equation}
q^\alpha H_{\alpha \beta}=q^\beta H_{\alpha \beta}=q^\alpha L_{\alpha
  \beta}=q^\beta L_{\alpha \beta}=0\,.
\end{equation}
The lepton tensor can thus be replaced by 
\begin{equation}
\tilde{L}_{\alpha \beta}={\rm Tr}(\not \! p_2' \gamma_\alpha \not \! q \not \!
p_2 \not \! q \gamma_\beta)\,,
\end{equation}
where terms proportional to the electron mass have been neglected, as they
are suppressed by $m_e/m_\mu$.  Straightforward evaluation then yields
\begin{equation}
\tilde{L}_{\alpha \beta}=-4q^2({p_2'}_\alpha {p_2}_\beta + {p_2}_\alpha
      {p_2'}_\beta -p_2' \cdot p_2 g_{\alpha \beta}) - 8 p_2' \cdot q p_2
      \cdot q g_{\alpha \beta}\,.
\end{equation}
The evaluation of $\overline{|{\cal M}|^2}$ reveals that effects due
to the Pauli form factor $F_2$ are suppressed by $Q^2/4 M^2$.  Below
pion-production threshold this parameter is at most 0.02, and so in
what follows we neglect the contribution to $H^{\alpha \beta}$ from
the nucleon Lorentz structure $i \sigma^{\mu \nu} q_\nu$. 
For the proton,
\begin{equation}
F_1(Q^2)=1 - \frac{1}{6} \langle r_p^2\rangle Q^2 + {\cal O}(Q^4),
\end{equation}
with $\langle r_p^2 \rangle^{1/2}=0.895(18)~{\rm fm}$~\cite{Sick03}, and so
$F_1(Q^2)=1$ up to a few per cent correction at the kinematics of
interest here. Under these approximations, we obtain
\begin{eqnarray}
H^{\alpha \beta}&=&{\rm Tr}((\not \! p' + M) \gamma^\alpha (\not \!  p +
M) \gamma^\beta)\\ &=&4({p'}^\alpha p^\beta + {p}'^\beta p^\alpha +
(M^2 - p' \cdot p) g^{\alpha \beta})\,.
\end{eqnarray}
Contraction of the tensors $H$ and $L$ then yields 
\begin{eqnarray}
\overline{|{\cal M}|^2}&=&\frac{16 e^4 v^2}{\Lambda^4} \frac{1}{(Q^2)^2} \{Q^2[2
(p_2' \cdot p') (p_2 \cdot p) + 2 (p_2' \cdot p) (p' \cdot p_2) - 2
M^2 (p_2' \cdot p_2)]\nonumber\\ && - 4 (p_2' \cdot q) (p_2 \cdot q) (p'
\cdot p) -8 M^2 (p_2' \cdot q) (p_2 \cdot q) + 8 (p' \cdot p) (p_2 \cdot
q) (p_2' \cdot q)\}\,.
\end{eqnarray}
Dropping terms which are suppressed by at least one power of $Q^2/M^2$
relative to the dominant contribution, we are left with
\begin{equation}
\overline{|{\cal M}|^2}=\frac{16 e^4 v^2}{\Lambda^4} \frac{1}{(Q^2)^2}
\left[(Q^2)(s-M^2-m_\mu^2 - Q^2)(s-M^2) -M^2(m_\mu^2 + Q^2)m_\mu^2 +
O(Q^6)\right]\,,
\end{equation}
where $s=(p+p_2)^2$.

Working now in the lab frame, and neglecting nucleon recoil, we have
\begin{eqnarray}
E_\mu&=&E_e;\\ q^2&\equiv&-Q^2=m_\mu^2 - 2 E_e^2 + 2 E_e \sqrt{E_e^2 -
m_\mu^2} \cos \theta_\mu\,,
\end{eqnarray}
where $\theta_\mu$ is the angle between the outgoing muon and the
incoming electron beam. The differential cross section is then
\begin{equation}
\frac{d \sigma}{d \Omega_\mu}=\frac{4 \alpha^2 v^2}{\Lambda^4} \sqrt{\bar{E_e}^2
- 1}~~f(\bar{E_e}, \cos \theta_\mu)\,,
\end{equation}
with $\bar{E_e}=E_e/m_\mu$ and
\begin{equation}
f(\bar{E_e}, \cos \theta_\mu)=\frac{8 \bar{E_e}^4 - 6 \bar{E_e}^2 + 2 \bar{E_e}
\sqrt{\bar{E_e}^2 - 1}(1 - 4 \bar{E_e}^2) \cos \theta_\mu} {(2 \bar{E_e}^2 - 2
\sqrt{\bar{E_e}^2 - 1} \bar{E_e} \cos \theta_\mu - 1)^2}\,.
\end{equation}
Integrating this over the muon solid angle $\Omega_\mu$, we obtain
\begin{equation}
\sigma=\frac{16 \pi \alpha^2 v^2}{\Lambda^4} \sqrt{\bar{E_e}^2 - 1}\left[2
\frac{\sqrt{\bar{E_e}^2 - 1}}{\bar{E_e}} \ln \left(\frac{\bar{E_e} +
\sqrt{\bar{E_e}^2 -1}}{\bar{E_e} - \sqrt{\bar{E_e}^2 -1}}\right)-1\right]\,.
\label{eq:BSMmuconvXSn}
\end{equation}
For $E_e$ just below pion threshold, the kinematic factor in the 
square brackets is
about 2. Taking $v \approx 200$ GeV and $\Lambda=1$ TeV results in a
predicted cross section on the order of $100$ fb, which would definitely be
observable. Including the nucleon-recoil terms neglected in the 
derivation of Eq.~(\ref{eq:BSMmuconvXSn}) would result in corrections of order
$\frac{E_e}{M}$, which are potentially as large as 20\% or so, but do not
change the order-of-magnitude of $\sigma$.

On the other hand, dimension-six BSM operators of the type in
Eq.~(\ref{eq:LBSM2}) do not induce effects mediated by low-momentum
photons. If such operators do not contain additional derivatives they
cannot produce powers of the nucleon mass in the numerator, and so the
largest cross section they can yield is
\begin{equation}
\sigma \sim \frac{\alpha^2 m_\mu^2}{\Lambda^4}\,,
\end{equation}
which is suppressed by $\left(\frac{m_\mu}{v}\right)^2 \simle
10^{-6}$ compared to the long-range mechanism depicted in
Fig.~\ref{susy}.  Operators $O$ that contain additional derivatives
will be suppressed even further by at least one factor of the small
parameter $\frac{M}{\Lambda}$.

The result (\ref{eq:BSMmuconvXSn}) suggests that electron scattering
from a proton target could provide access to BSM physics over a
sizeable range of $\Lambda$. However, this prediction does not take
into account the constraint on the $\mu \rightarrow e \gamma$ coupling
from the non-observance of this muon decay branch. As we shall see,
this places stringent limits on the size of the cross section for the
process in Fig.~\ref{susy}.

The operator in Eq.~(\ref{eq:LBSM1}) produces an amplitude for the
rare decay $\mu \rightarrow e \gamma$ that, in the muon rest frame,
takes the form
\begin{equation}
{\cal M}_{\mu \rightarrow e \gamma}=-\frac{i e v}{\Lambda^2} \bar{u}_e(-{\bf
q}) [\gamma^\alpha,\not \! q]u_\mu(0) \varepsilon_\alpha\,,
\end{equation}
where $\varepsilon$ is the photon polarization vector, and $q$ is the photon
four-momentum. From this, we obtain
\begin{equation}
\overline{|{\cal M}_{\mu \rightarrow e \gamma}|^2}=\frac{2 e^2 v^2
m_\mu^3}{\Lambda^4}\,.
\end{equation}
Converting this to a decay rate, and integrating over final electron states,
we find
\begin{equation}
\Gamma_{\mu \rightarrow e \gamma}=\frac{\alpha v^2 m_\mu^3}{2 \Lambda^4}\,.
\end{equation}
If we now use the result for the predominant muon decay
mode~\cite{Cheng:1985bj}
\begin{equation}
\Gamma_{\mu \rightarrow e \nu_\mu \bar{\nu}_e}=\frac{G_F^2 m_\mu^5}{192
  \pi^3}\,,
\end{equation}
(here $G_F=\frac{e^2}{2^{5/2}M_W^2{\rm sin}^2\theta_W}$ is the Fermi coupling
constant, with $M_W=80.41$ GeV the W-boson mass and $\theta_W$ the
Weinberg angle) we find that the branching ratio for $\mu \rightarrow
e\gamma$ is
\begin{equation}
{\rm BR}(\mu \rightarrow e \gamma)=96 \pi^3 \alpha \frac{v^2}{\Lambda^4}
\frac{1}{m_\mu^2 G_F^2}\,.
\label{eq:BRmuegamma}
\end{equation}
But the factor 
$\frac{v^2}{\Lambda^4}$ which appears here is the same as
that in the pre-factor in Eq.~(\ref{eq:BSMmuconvXSn}). Consequently,
we can eliminate this factor between Eqs.~(\ref{eq:BSMmuconvXSn}) and
(\ref{eq:BRmuegamma}) to obtain
\begin{equation}
\sigma \approx \frac{\alpha}{3 \pi^2} (m_\mu^2 G_F)^2 \frac{1}{m_\mu^2} {\rm BR}(\mu
  \rightarrow e \gamma)\,.
\end{equation}
Using ${\rm BR}(\mu \rightarrow e \gamma) < 4.9 \times
10^{-11}$~\cite{Yao:2006px},  we see that
\begin{equation}
\sigma < 7.0 \times 10^{-15}~{\rm fb}\,.
\label{eq:sigmaBSMnumber}
\end{equation}
This is a model-independent constraint on the contribution to the
cross section for $e p \rightarrow \mu p$ from photon exchange.
Interpreted as a bound on $\Lambda$, we find $\Lambda \geq 1.5 \times
10^4$ TeV.  A similar bound on $\sigma$ was derived within the context
of a MSSM calculation of the electron-nucleus to muon-nucleus cross
section and ${\rm BR}(\mu \rightarrow e \gamma)$ in
Ref.~\cite{Blazek:2004cg}. However, that number was seven orders of
magnitude larger than the result of
Eq.~(\ref{eq:sigmaBSMnumber}). Part of the difference arises from
Ref.~\cite{Blazek:2004cg}'s consideration of a target with $Z=70$.

One might ask why the apparent scale $\Lambda$ in Eq.~(\ref{eq:LBSM1})
is so large---or, equivalently, why the coupling is ``unnaturally''
small. One possible explanation arises in the scenario known as
``minimal-flavor violation''~\cite{Cirigliano:2005ck}. There the
physics beyond the Standard Model breaks the lepton-number symmetry of
the Standard Model in the same fashion in which it is broken by
neutrino mixing. This scenario can account for the small branching
ratio for $\mu \rightarrow e \gamma$ in a natural way as long as the
product of the relevant neutrino mass and the scale of lepton-flavor
violation $\Lambda$ is smaller than $v^2$.

Regardless of what physics determines ${\rm BR}(\mu \rightarrow e
\gamma)$, our calculations show that the bound on this quantity is
sufficiently stringent to preclude the observation of any $e p
\rightarrow \mu p$ cross section from the diagram of
Fig.~\ref{susy}. Indeed, the contribution of the operator in
Eq.~(\ref{eq:LBSM1}) is constrained so strongly by the non-observation
of this muon decay branch that effects from diagrams with short-range
operators of the form (\ref{eq:LBSM2}) are worth considering.  In
particular, if such EMU's involved a different scale
$\tilde{\Lambda}$, with $\tilde{\Lambda} \ll \Lambda$ of
Eq.~(\ref{eq:LBSM1}), they may produce a larger effect than
(\ref{eq:sigmaBSMnumber}). However, the estimates provided above
indicate that for $\tilde{\Lambda}=1$ TeV, contributions from these
dimension-six operators to the $e p \rightarrow \mu p$ cross section
would be at most $\sim 10^{-4}$ fb. The conclusion therefore is that
beyond-the-Standard-Model physics is unlikely to result in any
measurable production of muons when an electron beam impinges on a
proton target.


\section{Muon production via Standard-Model Electroweak Processes}

\label{sec-EMcrosssection}

\begin{figure}[t]
\includegraphics[height=5cm]{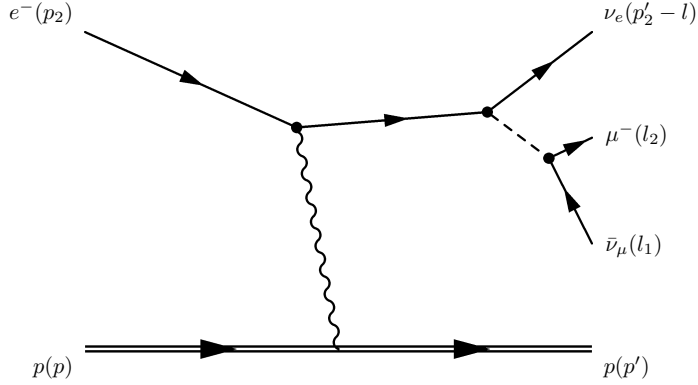}
\caption{\label{feyndiag}Leading-order Feynman diagram for the process
  $e^-p\rightarrow \mu^-\overline{\nu}_{\mu}\nu_ep$. Solid lines
  represent leptons, the double line is a proton, and the dashed line
  is a $W^-$ boson. The four-momentum carried by each external-state
  particle is indicated in parentheses.}
\end{figure}

In this section, we evaluate the scattering cross section for the
process $e^- p \rightarrow \mu^- \bar{\nu}_\mu \nu_e p$. The
dominant contribution to the scattering amplitude comes from
single-photon exchange given by the Feynman diagram in
Fig.~\ref{feyndiag}.
Applying the usual QED and Fermi-theory Feynman rules to the upper
vertices in the diagram of Fig.~\ref{feyndiag}, and using the
single-nucleon current of Eq.~(\ref{eq:nuccurrent}) for the
virtual-photon-nucleon vertex, we get
\beqy i{\cal M}&=&\left[-ie\ub_N({\bf p}')\left\{\gm F_1(Q^2) +
\kappa F_2(Q^2) \frac{i}{2M} \sigma^{\mu\rho}q_{\rho}\right\}u_N({\bf p})\right] \nonumber \\
&\times&\left(\frac{-ig_{\mu\nu}}{q^2}\right) \nonumber \\
&\times&\frac{G_F}{\sqrt{2}}\left[\ub_{\mu} ({\bf l}_2)\ga(1-\gf)v_{\bar{\nu}_\mu}({\bf l}_1)\right]
\left[\ub_e({\bf p}_2'-{\bf
    l})\gda(1-\gf)i\frac{\not \! p_2' +m_e}{{p_2'}^2-m_e^2}(-ie\gn)
u_e({\bf p}_2)\right]
\,,
\label{matelement}
\eeqy
\ni 
For unpolarized electrons, the
spin-summed-and-averaged squared matrix element can be expressed
as
\beqy \overline{|{\cal M}|^2}&=&\frac{c_{EW}^2(Q^2)}{2({p_2'}^2-m_e^2)^2}H^{\mu\nu}L^{\alpha\beta}
W_{\alpha\mu\nu\beta} \,;\nonumber \\
H^{\mu\nu}&=&\Tr\left[({\not \! p}'+M)\left\{\gm F_1 +
\frac{\kappa}{2M}i\sigma^{\mu\rho}q_{\rho}F_2\right\}(p\sla+M) \left\{\gn
F_1-\frac{\kappa}{2M}i\sigma^{\nu\theta}q_{\theta}F_2\right\} \right]\,,
\nonumber \\
L^{\alpha\beta}&=&\Tr\left[\not \! l_2\ga(1-\gf) \not \! l_1\gb(1-\gf)\right]\,, \nonumber
\\ W_{\alpha\mu\nu\beta}
&=&\Tr\left[(\not \! p_2'-l)\gamma_{\alpha}(1-\gf)(\not \! p_2'+m_e)
\gamma_{\mu}(\not \! p_2 +m_e)\gamma_{\nu}
(\not \! p_2'+m_e)\gamma_{\beta}(1-\gf)\right]\eeqy
\ni with $l \equiv l_1 + l_2$ and 
\beq
\label{cew}
c_{EW}^2(Q^2)=\frac{8\pi^2\alpha^2G_F^2}{Q^4}\,.
\eeq
The computation of the traces in $W_{\alpha\mu\nu\beta}$ is facilitated by the
Chisholm identity
\beq \gamma^{\alpha}\gamma^{\beta}\gamma^{\gamma} =
g^{\alpha\beta}\gamma^{\gamma} + g^{\beta\gamma}\gamma^{\alpha} -
g^{\alpha\gamma}\gamma^{\beta} +
i\epsilon^{\alpha\beta\gamma\delta}\gamma_{\delta}\gf \,. \eeq
Performing the contractions, we obtain
\beqy
\label{firstM}
\overline{|{\cal M}|^2}&=&\frac{c_{EW}^2}{{p_2'}^4}2^7
\left[\left(p_2'-l\right) \cdot l_1\right]
\left\{4F_1^2(A_1)-\left(\frac{\kappa}{2M}\right)^2F_2^2(A_2)-2\kappa
F_1F_2(A_{12})\right\}; \nonumber \\ (A_1) &=&
2(p_2'\cdot l_2)[(p\cdot p_2)(\pp\cdot p_2')+(\pp\cdot p_2)(p\cdot p_2')-M^2(p_2\cdot p_2')]\nonumber \\
&&-{p_2'}^2[(p\cdot p_2)(\pp\cdot l_2)+(\pp\cdot p_2)(p\cdot l_2)-M^2(p_2\cdot l_2)],\nonumber \\ (A_2) &=&
2(p_2'\cdot l_2)[2(P\cdot p_2)(P\cdot p_2')q^2+8M^2(q\cdot p_2)(q\cdot p_2')
+(p_2\cdot p_2')q^4]\nonumber \\
&&-{p_2'}^2[2(P\cdot p_2)(P\cdot l_2)q^2+8M^2(q\cdot p_2)(q\cdot l_2)+(p_2\cdot l_2)q^4], \nonumber \\
(A_{12}) &=&
2(p_2'\cdot l_2)[2(q\cdot p_2)(q\cdot p_2')+q^2(p_2\cdot p_2')]-{p_2'}^2[2(q\cdot p_2)(q\cdot l_2)+
q^2(p_2\cdot l_2)]
\,, \eeqy
\ni where $P=(p+\pp)$ and terms of ${\cal O}(m_e^2)$ have been dropped
as ${p_2'}^2 \gg m_e^2$. To eliminate the interference terms
involving $F_1F_2$, we reexpress $F_1$ and $F_2$ through the Sachs
form factors~\cite{RGS64}
\beq G_E=F_1+\frac{\kappa q^2}{4M^2}F_2 \quad {\rm and } \quad
G_M=F_1+\kappa F_2 \eeq
to get
\beqy
\label{finalM}
\overline{|{\cal M}|^2}&=&\frac{2^7c_{EW}^2[(p_2'-l)\cdot l_1]}{{p_2'}^2}
\biggl\{{\cal T}_1-2\frac{{p_2'}^2}{p_2'\cdot l_2}{\cal T}_2\biggr\}\quad;\nonumber\\
{\cal T}_1&=&G_{\tau}\left[P^2(p_2\cdot l_2)-2(P\cdot p_2)(P\cdot l_2)\right]+G_M^2
\left[2(q\cdot p_2)(q\cdot l_2)+(p_2\cdot l_2)q^2\right]\,,\nonumber \\
{\cal T}_2&=&G_{\tau}\left[P^2(p_2\cdot p_2')-2(P\cdot p_2)(P\cdot p_2')\right] +
G_M^2\left[2(q\cdot p_2)(q\cdot p_2') + (p_2\cdot p_2')q^2\right]\,,
\eeqy
\ni where $G_{\tau}=(G_E^2+\tau G_M^2)/(1+\tau)$ with $\tau=Q^2/4M^2$.  

In the laboratory frame, the differential cross section is given by
\beqy
\label{cross_sec}
d\sigma&=&\frac{1}{4ME_e} \int\frac{d^3l_1}{(2\pi)^32E_{\nu_{\mu}}}
\frac{d^3l_2}{(2\pi)^32E_{\mu^-}}\frac{d^3\pp}{(2\pi)^32E_{\pp}}
\frac{d^3l_{\nu_e}}{(2\pi)^32E_{\nu_e}}\overline{|{\cal M}|^2} \nonumber \\ &&\times(2\pi)^4\int
d^4p_2'\,\delta^{(4)}(p_2+p-p_2'-\pp) \int
d^4l\,\delta^{(4)}(l-l_1-l_2)\delta^{(4)}(p_2'-l-l_{\nu_e}) \,, \eeqy
\ni where $E_e$ is the energy of the incoming electron in the
laboratory frame, $l_{\nu_e}$ is the electron-neutrino's 4-momentum
and the last two delta functions and integrals have been inserted as unity to
simplify calculations. The steps given in
Appendix~\ref{ap-ewphasespace} then allow us to obtain from
Eq.~(\ref{cross_sec}) an expression for the differential
cross section per unit solid angle subtended by
the detected muon at fixed beam energy:
\beqy
\label{cross_sec4}
\frac{d\sigma}{d\Omega_{\mu}} &=& \frac{\alpha^2G_F^2}{32\pi^5M^3E_e}
\int_{m_{\mu}}^{E_e}dE_{\mu}\sqrt{E_{\mu}-m_{\mu}^2}
\int_{q_0^l}^{q_0^u}\frac{dq_0}{q_0^2}\sqrt{q_0^2+2Mq_0}
\int_{{\rm
cos}\theta_q^l}^{{\rm cos}\theta_q^u} \,d({\rm cos}\theta_q) \nonumber \\ 
&\times&\int_{\phi_q^l}^{\phi_q^u}d\phi_q \,(p_2-q-l_2)^2 \left[\frac{p_2'.l_2}{{p_2'}^4}{\cal T}_2
-\frac{1}{2{p_2'}^2}{\cal T}_1\right]_{(p_2'=(p_2-q),\, q^2=-2Mq_0)} \,.
\eeqy

The limits on the $q_0$ integral are determined by the electron beam
energy. The lower limit of the integral arises because energy transfer
to the proton without any momentum transfer is not possible in elastic
scattering: the target recoils.  In the limit that the muon is
produced at rest (${\bf l}_2=0$), an analytic expression for both the
upper and lower limit of the $q_0$ integration can be obtained. This can guide
intuition on the importance of collective excitations when the target
is replaced by a heavy nucleus (see Sec.~\ref{sec-nucleus}). We find
\beqy q_0^{\pm}&=&-\frac{B\mp\sqrt{B^2-4AC}}{2A} \,;\nonumber \\
A&=&4(E_e+M-m_{\mu})^2-4E_e^2\,;\quad C=m_{\mu}^2(2E_e-m_{\mu})^2
\,,\nonumber\\
B&=&2\left[2(E_e+M-m_{\mu})m_{\mu}(2E_e-m_{\mu})-4E_e^2M\right]\,.
\eeqy
Furthermore,
\beq
q_0^-\equiv \frac{m_{\mu}^2}{2(M-m_{\mu})}\leq E_e-E_{\mu}\,, 
\eeq
so that the minimum electron beam energy
$E_e^{\rm min}$ for muon production is then determined by requiring
$q_0^-=E_e^{\rm min}-m_{\mu}$, which yields
\beq E_e^{\rm min} = \frac{m_{\mu}(2M-m_{\mu})}{2(M-m_{\mu})}=111.6\,{\rm
MeV}. 
\eeq
The quantity $q_0^-$ is thus at least 6 MeV, whereas
\beqy
q_0^+=\frac{(2E_e-m_{\mu})^2}{2(2E_e-m_{\mu}+M)}
\eeqy
which is always less than $E_e-m_{\mu}$. The corresponding $Q^2$
ranges from 0.01 GeV$^2$ to a maximum of $2Mq_0^+$. The angle between
the electron and muon neutrinos (whose masses are neglected) is
constrained to $0\leq \angle({\bf l}_1,{\bf l}_{\nu_e})\leq \pi/2$ by
the step-function $\Theta(Q^2)$.  The maximum energy in neutrinos is
$E_{\nu}^{\rm max}=E_e-q_0^--m_{\mu}=E_e-E_e^{\rm min}$.

It is noteworthy that the differential cross section is independent of
the azimuthal  angle $\phi_{\mu}$. Dependence  on $\phi_{\mu}$ appears
explicitly  in the matrix  element through  the dot  products $(P\cdot
l_2)$ and  $(q\cdot l_2)$ as well  as implicitly in  the step function
$\Theta\left((k-q-l_2)^2\right)$, but  this dependence drops  out once
the  ${\rm d}\phi_q$  integration is  performed.  Only  differences of
azimuthal angles ($\phi_q-\phi_{\mu}$) appear in the ${\rm d}\phi_q$
integrand as well as in the limits  on this integral,  so the  integral remains
invariant.   Therefore  experiments  to   measure  this   process  are
characterized by $\theta_{\mu}$ alone, and the $\phi$-independence can
be used to  increase the total number of  counts (thereby decreasing
the statistical error) by  positioning several detectors in an annulus
at the same $\theta_{\mu}$.

The integrals in Eq.~(\ref{cross_sec4}) were evaluated numerically,
details of which are presented in Appendix~\ref{ap-numerical}.  The
result for the differential cross section $d\sigma/d\Omega_\mu$ as a
function of electron beam energy at a fixed value of
$\theta_{\mu}=\pi/3$ will be presented in
Sec.~\ref{sec-strongcrosssection}. The corresponding {\it total}
muon-production cross section rises from $3 \times 10^{-16}$ fb at
$E_e=120$ MeV to $4 \times 10^{-13}$ fb at $E_e=140$ MeV.

The cross section for the reaction $e^- p \rightarrow  \mu^- p \nu_e
\bar{\nu}_\mu$ is thus much smaller than
the low-energy approximation to the Rosenbluth
cross section for $ep$ elastic scattering~\cite{HM}:

\beq \frac{d\sigma}{d\Omega}\approx
\frac{\alpha^2F_1^2}{2E_e^2}\left[\frac{1}{{\rm sin}^2\theta{\rm
      tan}^2\theta}+{\cal
    O}\left(\frac{E_e^2}{M^2}\right)\right]. 
\label{eq:Rosenbluth}
\eeq At $E_e=140$ MeV
this is $\sim 10^7$ fb for all but forward angles where the Coulomb
singularity occurs. Thus, muon production via standard-model
electroweak processes is down by 20 orders of magnitude as compared to
elastic electron-proton scattering. Much of this suppression comes
from the extra factor $G_F^2l_{\nu_e}^2(l_{\mu}\cdot l_{\nu_{\mu}})$ (with
$l_x$ the four-momentum of lepton $x$). Numerically, $G_F^2\sim
10^{-10}$ GeV$^{-4}$, $l_{\nu_e}\sim 0.01$ GeV, $l_{\mu}\sim 0.1$ GeV,
and $l_{\nu_{\mu}}\sim 0.01$ GeV, and so this factor already implies a
suppression of 17 orders of magnitude.

Further suppression occurs due to the lower bound on the virtuality of
the exchanged photon, which was explained above, and persists even if
the muon is detected at $\theta_\mu=0$. The Coulomb divergence that is
manifest in Eq.~(\ref{eq:Rosenbluth}) as $\theta \rightarrow 0$ and is 
associated with $Q^2 \rightarrow 0$ does not
appear in $e^- p \rightarrow \mu^- p \nu_e \bar{\nu}_\mu$. The suppression
relative to the Rosenbluth cross section is therefore even more severe
than is implied by the dimensional analysis in the previous
paragraph. The lack of enhancement from the exchanged photon going
soft is an important feature of muon production.

The cross section predicted for standard-model electroweak $\mu^-$
production at the largest energy considered here, $E_e \approx m_\pi$,
is thus two orders of magnitude larger than the largest BSM cross
section predicted by Eq.~(\ref{eq:sigmaBSMnumber}). We limit the
indicent electron energy to
$E_e<m_{\pi}$ as for incident energies larger than the pion
mass, strong-interaction processes involving the production of an
on-shell pion which then decays to a muon will swamp the purely
electroweak diagram of Fig.~\ref{feyndiag}. In the following section, 
we show that a sub-threshold version of the pion-production process yields
muon-production cross sections that are significantly larger than
those obtained through the mechanism discussed in this section.


\section{Muon production via sub-threshold pion production}
\label{sec-strongcrosssection}

\begin{figure}[t]
\includegraphics[width=16cm]{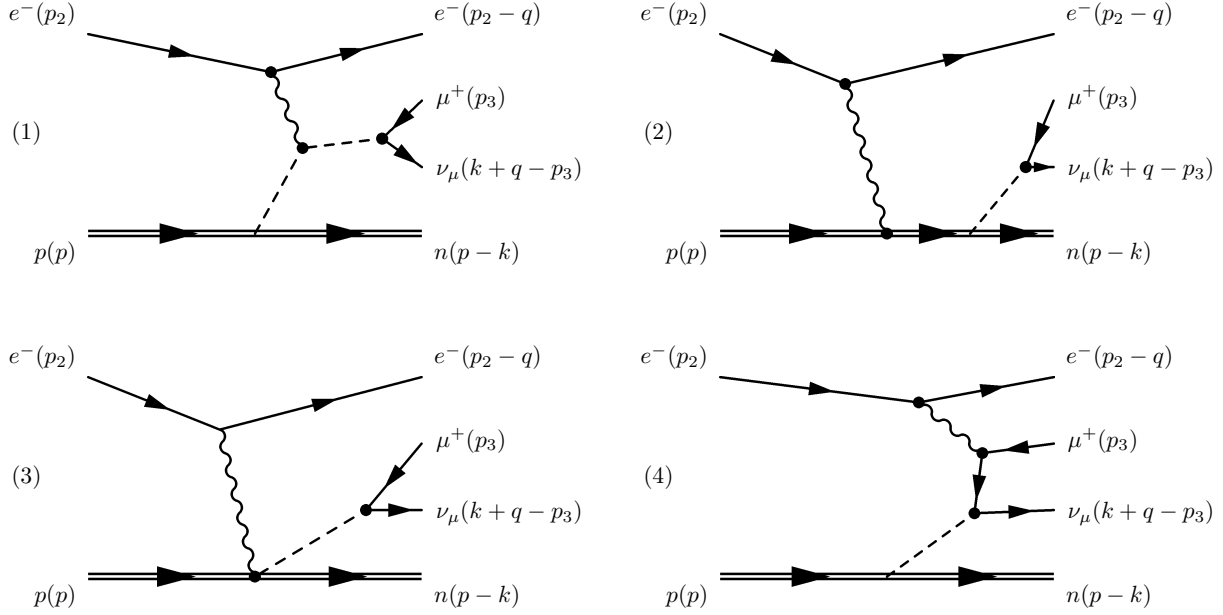}
\caption{\label{fig:chiptdiagrams} Leading-order $\chi$PT
  contributions to $\mu^+$-production in electron-proton
  scattering. The double line denotes the nucleon, the thin solid
  lines denote leptons and the dashed lines denote pions.  Particle
  momenta are indicated in parentheses. The numbers in parentheses
  correspond to the individual amplitudes computed below.}
\end{figure}

In electron-proton collisions, muons can also be generated through
processes in which a virtual photon couples to a pion which
then decays into a muon and a neutrino (Fig. \ref{fig:chiptdiagrams}).
This is a manifestation of the ``pion cloud'' of the nucleon, and the
processes depicted in Fig.~\ref{fig:chiptdiagrams} are possible even
when the pion in the intermediate state is virtual, i.e. $E_e$ is
significantly below the pion mass.
Whereas only negatively charged muons can be produced in the processes
considered thus far, this strong-interaction process 
yields positively charged muons, together with a neutron in the final
state. In this section, we evaluate the differential cross section $\frac{d
  \sigma}{d \Omega_\mu}$ for the reaction: $e p \rightarrow e n \mu^+
\nu_\mu$.

Our calculations are performed using chiral perturbation theory
($\chi$PT), the low-energy effective field theory of QCD. $\chi$PT
incorporates QCD's (broken) $SU(2)_L \times SU(2)_R$ symmetry as well
as the pattern of chiral-symmetry breaking in QCD (for a recent review
see Ref.~\cite{Bernard:2006}).  Reactions involving pions, nucleons,
and photons (either real or virtual) can be straightforwardly and
systematically evaluated using $\chi$PT, as long as the energies
involved are well below the excitation energy of the
$\Delta$(1232). Here, we perform a tree-level calculation of the
process of interest using the leading-order $\chi$PT Lagrangian.  For
diagrams (1) to (3) in Fig.~\ref{fig:chiptdiagrams}
our calculation is equivalent to evaluating
the amplitude for charged-pion electroproduction at leading order
${\cal O}(e)$. Such a leading-order calculation is known to give a reasonable
description of the available data for charged-pion photoproduction
near threshold~\cite{Fearing:1999}.

The leading-order $\chi$PT Lagrangian describing the interactions between
pions, photons, and nucleons is given by \cite{Donoghue:1992dd}
\begin{equation}
{\cal L}_{\chi \rm{PT}}={\cal L}_{N\pi}^{(1)}+{\cal L}_{\pi\pi}^{(2)}~.
\end{equation}
Here ${\cal L}_{\pi\pi}^{(2)}$ denotes the leading-order
Goldstone-boson Lagrangian
\begin{equation}
\label{eq:Lagr_PIPI}
{\cal L}_{\pi \pi}=\frac{f_\pi^2}{4}\mbox{Tr}[D_\mu U (D^\mu U)^\dagger]
+\frac{f_\pi^2}{4}\mbox{Tr}(\chi U^\dagger + U\chi^\dagger)~,
\end{equation}
where to leading order in quark masses the matrix $\chi$ is $m_\pi^2$
times the identity matrix, 
and ${\cal L}_{N\pi}^{(1)}$ denotes the lowest-order Lagrangian involving
baryons:
\begin{equation}
\label{eq:Lagr_PI_N}
{\cal L}^{(1)}_{\pi N}= \bar{\Psi}\left(iD\hspace{-.6em}/
-\stackrel{\circ}{M}
+\frac{\stackrel{\circ}{g}_A}{2}\gamma^\mu \gamma_5 u_\mu\right)\Psi~.
\end{equation}
In the above equations,  
the pion fields are collected in the matrix 
$U=\exp(i\tau\cdot\pi/f_\pi)$, whereas the fields 
$u=\exp(i {\protect \mathbf \tau}\cdot{\pi}/(2f_\pi))$ and
$u_\mu=i(u^\dagger\partial_\mu u-u\partial_\mu
u^\dagger)$.  The quantities 
$\stackrel{\circ}{g}_A$ and $\stackrel{\circ}{M}$ denote
the axial coupling constant and the nucleon mass, respectively, 
in the chiral limit.  The
covariant derivative acting on the pion matrix is defined as
\begin{equation}
\label{eq:cov_deriv_pion}
D_\mu U\equiv\partial_\mu U -i r_\mu U+iU l_\mu~,
\end{equation}
where $r_\mu$ and $l_\mu$ denote the appropriate external fields and
external electromagnetic fields $\mathcal{A}_\mu$ are coupled to the pion field
by setting $r_\mu=l_\mu=\mathcal{A}_\mu$. The (chiral and $U(1)_{\rm em}$)
covariant derivative acting on the nucleon field is then
\begin{equation}
D_\mu\Psi=\biggl(\partial_\mu+\frac{1}{2}
\bigl(u^\dagger(\partial_\mu-ir_\mu)u + 
u(\partial_\mu-il_\mu)u^\dagger\bigr)-i(r_\mu+l_\mu)\biggr)\Psi \,.
\end{equation} 
As our leading-order computation involves only tree-level diagrams,  
we can employ the relativistic Lagrangian for nucleon and pion fields
without concern about contributions from loop graphs
that might violate power counting~\cite{GasserSainioSvarc}. We
will therefore use relativistic Feynman rules in what follows.
The amplitude corresponding to each of the diagrams contributing to 
$\mu^+$-production at leading order (see Fig.~\ref{fig:chiptdiagrams})
is then 
\bea
\nonumber
\mathcal{M}_1&=&-i\sqrt{2}\,V_{ud}G_F f_\pi g_{\pi NN} \,e^2
\frac{1}{(k+q)^2-m_\pi^2}\frac{1}{k^2-m_\pi^2}\frac{1}{q^2}\\
\nonumber
&&\qquad\times\bar{u}_\nu({\bf k+ q - p_3})
(k\hspace{-1.9mm}\slash+q\hspace{-1.8mm}\slash)(1-\gamma^5)v_\mu({\bf p_3})\\
\nonumber&&\qquad\times
\bar{u}_e({\bf p_2-q})(2k\hspace{-1.9mm}\slash+q\hspace{-1.8mm}\slash)u_e({\bf p_2})\\
&&\qquad\times\bar{u}_N({\bf p-k})\gamma^5 u_N({\bf p})~,
\eea
\bea
\nonumber
\mathcal{M}_2&=&-i\sqrt{2}\,V_{ud}G_F f_\pi g_{\pi NN} \,e^2
\frac{1}{(k+q)^2-m_\pi^2}\frac{1}{(p+q)^2-M^2}\frac{1}{q^2}\\
\nonumber
&&\qquad\times\bar{u}_\nu({\bf k+q-p_3})
(k\hspace{-1.9mm}\slash+q\hspace{-1.8mm}\slash)
(1-\gamma^5)v_\mu({\bf p_3})\\
\nonumber&&\qquad\times
\bar{u}_e({\bf p_2-q})\,\gamma_\mu\,u_e({\bf p_2})\\
&&\qquad\times\bar{u}_N({\bf p-k})\gamma^5
(p\hspace{-3mm}\slash+q\hspace{-1.8mm}\slash+M)\gamma^\mu u_N({\bf p})~,
\eea
\bea
\nonumber
\mathcal{M}_3&=&-i\sqrt{2}\,V_{ud}G_F f_\pi g_{\pi NN} \,e^2\frac{1}{2 M}
\frac{1}{(k+q)^2-m_\pi^2}\frac{1}{q^2}\\
\nonumber
&&\qquad\times\bar{u}_\nu({\bf k+q-p_3})
(k\hspace{-1.9mm}\slash+q\hspace{-1.8mm}\slash)
(1-\gamma^5)v_\mu({\bf p_3})\\
\nonumber&&\qquad\times
\bar{u}_e({\bf p_2-q})\,\gamma^\mu\,u_e({\bf p_2})\\
&&\qquad\times\bar{u}_N({\bf p-k})\,\gamma^5 \gamma_\mu\,u_N({\bf p})~,
\eea
\bea 
\nonumber \mathcal{M}_4&=&-i\sqrt{2}\,V_{ud}G_F f_\pi g_{\pi NN}
\,e^2 \frac{1}{(p_3-q)^2-m_\mu^2}\frac{1}{k^2-m_\pi^2}\frac{1}{q^2}\\
\nonumber &&\qquad\times\bar{u}_\nu({\bf k+q-p_3})
(k\hspace{-1.9mm}\slash+q\hspace{-1.8mm}\slash)
(1-\gamma^5)(p_3\hspace{-3mm}\slash-q\hspace{-1.8mm}\slash+m_\mu)
\gamma_\mu\,v_\mu({\bf p_3})\\ \nonumber&&\qquad\times
\bar{u}_e({\bf p_2-q})\,\gamma^\mu\,u_e({\bf p_2})\\
&&\qquad\times\bar{u}_N({\bf p-k})\gamma^5 u_N({\bf p})~,
\eea 
where $g_{\pi NN}=\frac{M g_A}{f_\pi}$ at this order, and we adopt
$g_A=1.26$, $f_\pi=92$ MeV, $M=939$ MeV.  Note that the crossed
counterpart of diagram (2) is zero if only leading-order couplings are
considered, as this process involves a neutron in the final state.

\begin{figure}
\bce \includegraphics[scale=0.5]{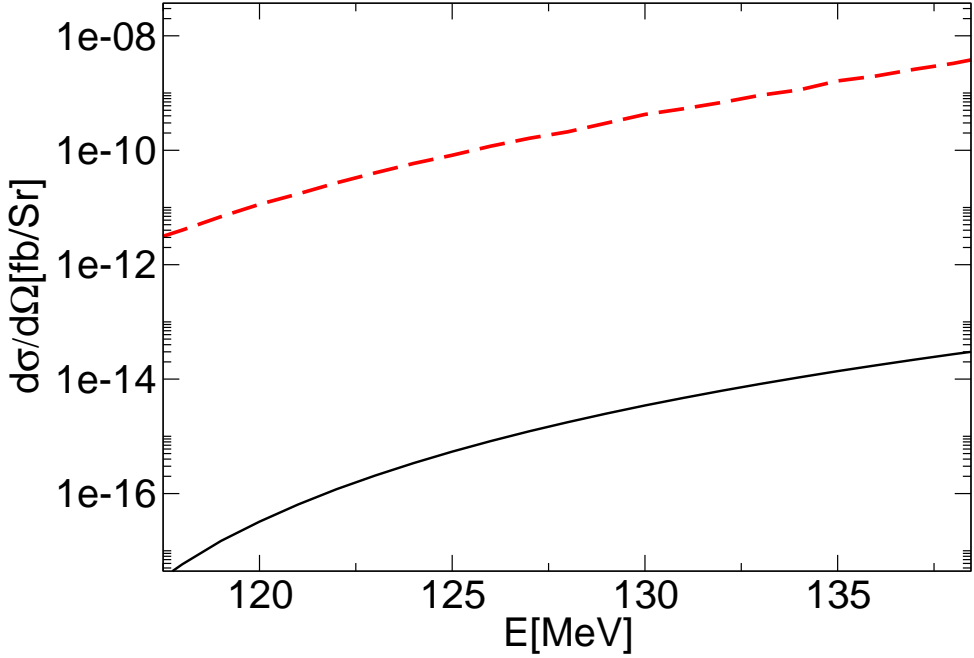} \ece
\caption{\label{crossfig}The differential cross section in femtobarns
per steradian versus the energy of the incident electron beam. The
solid line gives the result for production of a $\mu^-$ via the
process $e^- p \rightarrow \mu^- p \bar{\nu}_\mu \nu_e$ (see
 Sec.~\ref{sec-EMcrosssection}), and the
dashed line is the result for production of a $\mu^+$ via the
reaction $e^- p \rightarrow e^- n \mu^+ \nu_\mu$ (see
Sec.~\ref{sec-strongcrosssection}). 
Both results were evaluated for a representative muon angle
$\theta_\mu=\pi/3$.  }
\end{figure}

We evaluate the matrix elements ${\cal M}_1$--${\cal M}_4$ using the
package FeynCalc~\cite{Mertig:1990an}. This produces an expression for
the spin-averaged-and-summed squared matrix element $\overline{|{\cal
    M}|^2}$ that is lengthy and not particularly illuminating.  The
differential cross section is then
\begin{equation}
\frac{d \sigma}{d \Omega_\mu}=\frac{1}{2 M \,2 E_2}\int
\frac{\hbox{d}^3 p'\,\hbox{d}^3 p_2'\,\hbox{d}^3 p_3\,\hbox{d}^3 p_\nu
}{(2\pi)^{12}\,2 E_1' \,2 E_2' \,2 E_3 \,2 E_\nu}(2\pi)^4
\delta(p'+p_2'+p_3+p_\nu-p_2-p)\:\overline{|\mathcal{M}|^2}~,
\label{eq:pionphasespace}
\end{equation}
where $p_\nu=(E_\nu,{\bf p}_\nu)$ is the four-momentum of the
outgoing neutrino, and the four vectors $p'=p-k$, $p_2'=p_2-q$, and
$p_3$ (which is the outgoing muon momentum) are written in a similar fashion.
We evaluate the integrals in Eq.~(\ref{eq:pionphasespace}) by Monte
Carlo integration, obtaining a result that is numerically stable to
better than 5\% accuracy.

The results of our calculation are shown by the dashed line in
Fig.~\ref{crossfig}, where the energy dependence of the differential
cross section at a representative outgoing muon angle of
$\theta_{\mu}=\pi/3$ is displayed. The solid curve in this figure
shows the differential cross section for the production of $\mu^-$
through the photon and $W^-$ mediated mechanism of the previous
section.  The dashed curve shows results for the production of
$\mu^+$'s through virtual-photon exchange discussed in this section.
The differential cross section for $\mu^+$ production is four to five
orders of magnitudes larger than that for $\mu^-$ production.  Even
the $\mu^+$-production cross section is, however, very small: of order
$10^{-9}$ fb at the largest energy considered ($E_e=140$ MeV).  The
variation of the cross section with the angle $\theta_\mu$ is one
order of magnitude for both cross sections, so we predict a total
cross section for $\mu^+$ production of order $10^{-8}$ fb just below
the pion threshold.  

\begin{figure}
\bce \includegraphics[scale=0.5]{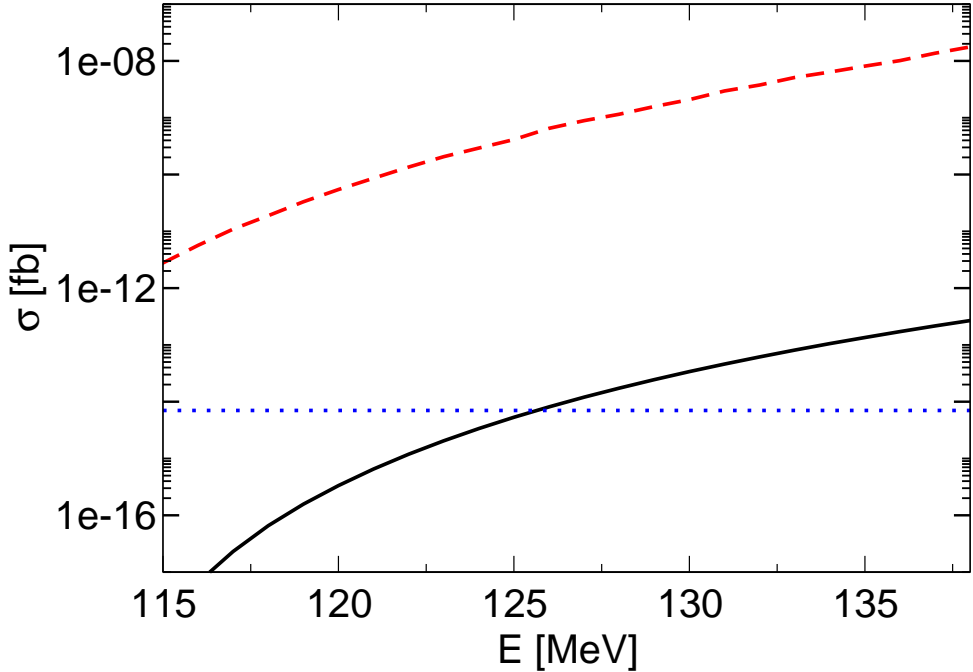} \ece
\caption{\label{fig-totalXSn}Total cross sections in femtobarns
  plotted against the energy of the incident electron beam. The solid
  line gives the result for production of a $\mu^-$ via the process
  $e^- p \rightarrow \mu^- p \bar{\nu}_\mu \nu_e$ (see
  Sec.~\ref{sec-EMcrosssection}), and the dashed line is the result
  for production of a $\mu^+$ via the reaction $e^- p \rightarrow e^-
  n \mu^+ \nu_\mu$ (see Sec.~\ref{sec-strongcrosssection}). The dotted
  line is the bound on BSM contributions obtained for $ep \rightarrow
  \mu p $ in Section~\ref{sec-BSM} by considering the dimension-five
  BSM operator and the non-observation of the decay $\mu \rightarrow e
  \gamma$.}
\end{figure}

The dependence on energy of the total cross section for 
the processes $e^- p \rightarrow e^- n \mu^+ \nu_\mu$ and $e^- p
  \rightarrow \mu^- p \bar{\nu}_\mu \nu_e$ (see 
  Sec.~\ref{sec-EMcrosssection}) is shown in
  Fig.~\ref{fig-totalXSn}. Also shown in Fig.~\ref{fig-totalXSn} is
  the bound of Eq.~(\ref{eq:sigmaBSMnumber}) for $ep \rightarrow \mu p $ from
  photon exchange.  Note that in Fig.~\ref{fig-totalXSn} we do not
  display results for energies exceeding 140 MeV, because above that
  energy the $n \pi^+$ channel opens and $\mu^+$'s are copiously
  produced through the decay of real pions.

Even at $E_e=140$ MeV, the cross section for $\mu^+$ production via
strong interactions is many orders of magnitude larger than the cross
section for BSM muon conversion in
Eq.~(\ref{eq:sigmaBSMnumber}). Indeed, it may be competitive with BSM
mechanisms even if dimension-six BSM operators that induce EMU are not
suppressed by, e.g. minimal lepton-flavor violation. Thus, any
experiment that searches for EMU on a proton target via BSM processes
should discriminate between the desired reaction and the channel $e p
\rightarrow e n \mu^+ \nu_\mu$. Such a discrimination requires either
detecting the outgoing electron, or detecting the charge of the
final-state muon.


\section{Muon production in electron-nucleus scattering}
\label{sec-nucleus}

If the proton target were replaced by a neutron target (e.g. via the
use of neutrons bound inside a deuterium nucleus), the incoming
electron can also interact with the neutron through its magnetic
moment. From Eq.~(\ref{matelement}), the matrix element for
magnetic-moment interactions introduces an extra factor---relative to
the dominant charge interaction---of $q_{\mu}/M$. This translates to a factor
$Q^2/M^2\approx q_0/M\approx 0.01$ in the cross section. Thus,
the electroweak process $e n \rightarrow \mu^-\nu_e\bar \nu_\mu n$  
yields a smaller muon-production
cross section than in the case of a proton.
In contrast, the process $e p \rightarrow e n \mu^+ \nu_\mu$
discussed in Sec.~\ref{sec-strongcrosssection} is associated with an
isovector matrix element at leading order in $\chi$PT, and so the
cross section for production of muons will be as large for a neutron
target as for a proton target. In this case the reaction is, however,
$e n \rightarrow e p \mu^- \bar{\nu}_\mu$. For neutrons this
muon-production reaction provides a signal that cannot be
distinguished from BSM electron-muon conversion by detection of the
charge of the final-state muon.  (As an aside, we note that in the
case of $Z^0$ exchange, scattering off a neutron is more favorable due
to its much larger weak charge as compared to the proton. However, the
appearance of an extra factor of $G_F$ renders the cross section due
to electromagnetic-weak interference terms negligible when compared to
photon exchange.)

The cross section for muon production is enhanced when the electron
scatters off a heavy nucleus. As an illustrative example, and to
estimate the expected enhancement over the nucleonic case, we consider
electron scattering on a lead nucleus ($^{208}{\rm Pb}$). The low
energy and low three-momentum transfer region is probed in
conventional nuclear spectroscopy. In this region, the elastic peak
appears first, although at $q_0^{\rm Pb}=-q^2/2m_{\rm Pb}$ instead of
$q_0^{\rm prot.}=-q^2/2M$, so that the exchanged photon appears to be
two orders of magnitude softer. In fact, though, the requirement of
producing the muon implies that the photon's virtuality is unchanged
from $Q^2\sim m_{\mu}^2$; as the differential cross section
scales as $\sim 1/(Q^2)^2=1/(2m_{\rm Pb}q_0^{\rm Pb})^2$ considering
only the effect of the heavier target yields no particular
enhancement. This fact can also be verified by counting powers of the
target mass and energy transfer in the expression for the differential
cross section in Eq. (\ref{cross_sec4}).

The large charge of lead ($Z=82$) does tend to increase the cross
section, although nuclear elastic form factors offset this effect
substantially. The typical three-momentum transfer involved in elastic
scattering is $|{\bf q}|=\sqrt{q_0^2+2m_{\rm Pb}q_0}\approx m_{\mu}$,
which corresponds to a spatial resolution of about 2 fm. But $R_{\rm
  Pb}\sim 7$ fm is the typical size of the charge distribution in
$^{208}{\rm Pb}$ as determined by fitting a conventional 2-parameter
Fermi distribution for a spherical nucleus~\cite{Bellicard}, so we do
not expect the lead nucleus to respond coherently to the
electromagnetic probe. This can be quantified if we approximate the
elastic form factor by the diffraction pattern from a spherical charge
distribution of radius $R_{\rm Pb}$. In so doing we obtain an overall
factor relative to the proton case of:
\begin{equation}
Z^2 F^2(|{\bf q}|)\approx Z^2 \left[\frac{3j_1(|{\bf q}|R_{\rm Pb})}{|{\bf q}|R_{\rm
Pb}}\right]^2 \,,
\end{equation}
where $j_1$ is the spherical Bessel function of the first kind. For
lead, $F^2\sim 10^{-2}$ at the $Q^2$'s of interest here, in good
agreement with form factors extracted from data on elastic scattering
from $^{208}{\rm Pb}$ in this region of energy and momentum
transfer~\cite{Frois}.  Therefore, if we replace the target proton by
a lead nucleus, we expect an overall increase of the cross section by
$Z^2F^2\sim {82}^2\times 10^{-2}\approx 67$.

Elastic scattering is not the complete story, however, because the
maximum energy of the exchanged photon is $\sim (E_e-m_{\mu})\sim 30$
MeV, which is sufficient to excite a tower of collective
states. Studies of inelastic form factors of the first few excited
states (for example, the $3^-$ octupole in $^{208}$Pb) reveal a
suppression of $\sim 10^{-2}$ or more compared to the elastic
peak~\cite{Kendall,Zeigler}. The width of these excited states is also
small (0.1 MeV for the $3^-$ state), therefore, at low energies of the
exchanged photon, the contribution of the elastic peak is dominant. At
slightly higher energies, $q_0\gtrsim 10$ MeV, giant monopole and
multipole resonances can be excited. These resonances are of empirical
importance in studies of nuclei as they carry non-zero isospin. Although
the resonances have large widths (1--5 MeV), their contribution to the
cross section will also be smaller than that from the elastic peak.

Finally, we inquire whether quasi-elastic scattering should be taken
into account.  By quasi-elastic scattering, we are referring to those
events in which a muon is produced {\it and} a nucleon is knocked out of
the nucleus.  This phenomenon requires the additional kinematic
restriction that the three-momentum transfer exceed the Fermi momentum
of the nucleon in the nucleus. Therefore, the relevant energy regime
is now defined by the conditions $\Theta(|{\bf q}|-k_F)$ and the theta
functions imposed above, i.e, $\Theta(E_e-q_0-E_{\mu})$ and
$\Theta((k-q-l_2)^2)$.  These two theta functions are unchanged from
the nucleonic case as they originate from the kinematics of the
leptonic portion of the process, which is unaffected by changing the
target from a nucleon to a nucleus. These restrictions imply that the
maximum value of $(k-q-l_2)^2$ is given by
\beq (E_e-q_0)^2-E_e^2-k_F^2+m_{\mu}^2-2m_{\mu}(E_e-q_0)+2E_ek_F{\rm
cos}(\hat{k}\hat{q})\geq 0 \,, 
\eeq
where the inequality imposed by the theta function $\Theta((k-q-l_2)^2)$ 
is satisfied so long as $0<q_0<q_0^<$, where
$q_0^<$ is the lesser root of the above quadratic in $q_0$. Clearly,
this requires that $q_0^<>0$, which is equivalent to the condition
\beq {\rm
cos}(\hat{k}\hat{q})\geq\frac{m_{\mu}}{k_F}+\frac{k_F^2-m_{\mu}^2}{2E_ek_F}
\,.
\eeq
As $|{\rm cos}(\hat{k}\hat{q})|\leq 1$, we obtain the restriction
\beq E_e\geq\left(\frac{k_F+m_{\mu}}{2}\right) \,.
\label{qelimit}
\eeq
If we assume a simple picture of the nucleus with constant density
$\rho\approx\rho_{\rm nuc}=0.16$ fm$^{-3}$, then $k_F\approx 260$ MeV,
which implies that $E_e\geq 187$ MeV.  This exceeds the
pion-production threshold in ordinary electron-nucleus scattering (no
muon production).  It is highly desirable that the electron beam
energy not be above the pion threshold at around 140 MeV, and in this case
we need not include the contribution from quasi-elastic scattering,
since Eq.~(\ref{qelimit}) makes clear that it is important only at energies
well above pion threshold. This is significantly different to the
usual situation in inelastic electron-nucleus scattering (i.e. without
muon production), in which pion production occurs at energies that
exceed the quasielastic peak. When muon production happens, additional
kinematic restrictions (viz., the energy cost of producing a muon)
imply that the quasielastic peak is only important at energies that
exceed the threshold for pion production. This is another
distinguishing feature of the muon-production process.


\section{Conclusions}
\label{sec-conclusion}

We have examined the possibility of discovering physics beyond the
Standard Model through lepton-flavor violation in fixed-target
electron scattering.  Our main findings can be summarized as:
\begin{itemize}
\item 
We have obtained a model-independent constraint on the magnitude of
LFV in electron-nucleon scattering from beyond-the-Standard-Model
effects using a general low-energy effective interaction with
couplings constrained by experimental bounds on the nonobservance of
$\mu \rightarrow e\gamma$.  The cross section for LFV from the
lowest-dimension operator, $\sigma < 7 \times 10^{-15}$ fb, is too
small to be experimentally accessible with current technologies.  The
contribution of higher-dimension LFV operators to $ep\rightarrow \mu
p$ could be larger, but is still unobservably small at present.  This
is in accord with similar estimates that have been made previously
within specific extensions of the Standard
Model~\cite{Diener:2004kq,Blazek:2004cg}.

\item We have identified two main sources of background in inclusive
  $ep$ scattering within the Standard Model when only the energy of
  the outgoing muon is measured, and performed detailed calculations
  of the relevant cross sections. The reaction $e^- p \rightarrow
  \mu^-\nu_e\bar\nu_\mu p$ is the principal background if the charge
  of the muon is measured, and its cross section varies from the order
  of $10^{-16}$ fb at incident electron energy $E_e = 120$ MeV to
  $10^{-13}$ fb at $E_e=140$ MeV.

\item If the charge of the muon is not measured, the dominant source
  of background comes from $\mu^+$s produced by the decay of virtual
  pions. Leading-order chiral perturbation theory gives this
  reaction's total cross section $\sigma(e p \rightarrow e n
  \mu^+\nu_\mu)$ to be about $10^{-11}$ fb at incident electron
  energy $E_e = 120$ MeV and $\sim 10^{-8}$ fb near the pion
  threshold. This background swamps any LFV
  signal in $ep$ scattering unless the outgoing electron is also
  detected or/and $\mu^+$ events are vetoed.

\item Using a heavy nucleus as a target enhances both the desired LFV
  effects and the background. At the low energies carried by the
  exchanged photon in $e^- p \rightarrow \mu^- p \nu_e\bar{\nu}_\mu$ the
  role of collective nuclear excitations can be neglected in
  comparison to the leading effects of elastic scattering from a
  finite-size target. This could enhance the cross section for $e^- p
  \rightarrow \mu^- p \nu_e \bar{\nu}_\mu$ by as much as two orders of
  magnitude, but the cross section is still
  too small to be experimentally detectable at present.
\end{itemize}


\begin{appendix}
\section{Phase-space evaluation for final state $p \mu \bar{\nu_\mu}
  \nu_e$}

\label{ap-ewphasespace}

In this appendix, we explain how to obtain Eq.~(\ref{cross_sec4}) from
Eq.~(\ref{cross_sec}). We first employ a useful relation for elastic
scattering
\beq \int\frac{d^3\pp}{2E_{\pp}}\delta^{(4)}(p+q-\pp) =
\frac{1}{2M}\delta\left(\frac{q^2}{2M}+q_0\right) \quad {\rm where} \quad
q_0=p_{20}^{'}-p_{20}\quad \,, \eeq
\ni 
which enables Eq.~(\ref{cross_sec}) to be rewritten as
\beqy
\label{cross_sec2}
d\sigma &=& \frac{2\pi}{8M^2E_e}\int\frac{d^3l_2}{(2\pi)^32E_{\mu^-}}\int
d^4q \int d^4p_2'\delta^{(4)}(p_2-p_2'-q)\delta\left(\frac{q^2}{2M}+q_0\right)
\nonumber \\ &\times& \int\frac{d^3l_1}{(2\pi)^32E_{\nu_{\mu}}}
\frac{d^3l_{\nu_e}}{(2\pi)^32E_{\nu_e}}\delta^{(4)}(p_2'-l_1-l_{\nu_e}-l_2)\overline{|{\cal M}|^2}
\,.  \eeqy
For massless neutrinos, the phase-space integrals over neutrino
momenta in the second line of Eq.~(\ref{cross_sec2}) can be rewritten
as
\beq
\label{twoint}
\int d^4L\,\delta^{(4)}(L-p_2'+l_2)\int\frac{d^3l_1}{(2\pi)^32E_{\nu_{\mu}}}
\frac{d^3l_{\nu_e}}{(2\pi)^32E_{\nu_e}}\delta^{(4)}(L-l_1-l_{\nu_e})\overline{|{\cal M}|^2} \,.
\eeq
\ni Noting from Eq.~(\ref{finalM}) that Eq.~(\ref{twoint}) has a factor
$(p_2'-l)\cdot l_1=l_{\nu_e}\cdot l_1$,
the integrals over $d^3l_{\nu_e}$ and $d^3l_1$ are~\cite{Jaikumar:2001hq}
\beqy \int\frac{d^3l_1}{(2\pi)^32E_{\nu_{\mu}}}
\frac{d^3l_{\nu_e}}{(2\pi)^32E_{\nu_e}}\delta^{(4)}(L-l_1-l_{\nu_e})~l_{\nu_e}.l_1
&=& \frac{\pi L^2}{4(2\pi)^6}\theta(L_0)\theta(L^2) \,, \eeqy
\ni where $L^2=L_0^2-{\bf L}^2$. Using the resultant expression for
Eq.~(\ref{twoint}) in Eq.~(\ref{cross_sec2}), performing the $d^4L$ and
$d^4p_2'$ integrations with the aid of corresponding delta functions, and using
Eq.~(\ref{cew}), we obtain
\beqy
\label{cross_sec3}
d\sigma &=& \frac{\alpha^2G_F^2}{4\pi^2M^2E_e}
\int\frac{d^3l_2}{2E_{\mu^-}}\int \frac{d^4q}{q^4}\delta
\left(\frac{q^2}{2M}+q_0\right){\cal I}(q,l_2)\,;\nonumber \\
{\cal I}(q,l_2)&=&(p_2-q-l_2)^2\left[\frac{p_2'.l_2}{{p_2'}^4}{\cal
    T}_2-\frac{1}{2{p_2'}^2}{\cal T}_1\right]_{p_2'=(p_2-q)}\hskip
-0.55in\Theta\left((p_2-q-l_2)^2\right)\Theta(E-E_{\mu^-}-q_0)\,.  \eeqy
where ${\cal T}_1,{\cal T}_2$ are given by Eq.~(\ref{finalM}).With the
aid of the only remaining delta function, the $\int d^4q$ can be
recast as
\beqy 
&&\int\frac{d^4q}{q^4}\delta
\left(\frac{q^2}{2M}+q_0\right){\cal I}(q,l_2)=\nonumber\\
&&M\int\frac{dq_0}{(-2q_0M)^2}
\sqrt{q_0^2+2Mq_0}\,\Theta(E_e-E_{\mu^-}-q_0) \nonumber \\ 
&\times&\int d\Omega_q\Theta\left((p_2-q-l_2)^2\right) (p_2-q-l_2)^2
\left[\frac{p_2'.l_2}{{p_2'}^4}{\cal T}_2-\frac{1}{2{p_2'}^2}{\cal T}_1\right]_{(p_2'=(p_2-q),
 q^2=-2Mq_0)} \,.  \eeqy
The step functions $\Theta(E-q_0-E_{\mu^-})$ and
$\Theta\left((p_2-q-l_2)^2\right)$ provide the upper and lower limits
on the $dq_0$ integral. The latter step-function also provides bounds
on the angular integrations involving $d{\rm cos}\theta_q,
d\phi_q$. This determines the support for the various integrals as
[$q_0^l, q_0^u$], [${\rm cos}\theta_q^l, {\rm cos}\theta_q^u$],
[$\phi_q^l,\phi_q^u$] and leads to (\ref{cross_sec4}).  In our
numerical calculations, we have used a constant value for
$G_F(Q^2=0.01{\rm GeV}^2)$=$1.05\times 10^{-5}$GeV$^{-2}$ as
determined by its Standard-Model running in the $\overline{MS}$
scheme~\cite{Czarnecki:2000ic}.
\section{Numerical notes}

\label{ap-numerical}

The integrals in Eq.~{(\ref{cross_sec4})} are performed as follows.
We choose the $+\hat{z}$ axis to be along the electron beam
direction. The polar angle $\theta_{\mu}$ is measured from the
$+\hat{z}$-axis in the vertical plane containing this axis. The
azimuthal angle $\phi_{\mu}$ is measured anti-clockwise from the
(arbitrary) $-\hat{z}$ axis in a plane containing this axis.
The position of the
detected muon is then uniquely specified by the angles $\theta_{\mu}$
and $\phi_{\mu}$.  Once the position of the muon is specified as
above, for a fixed momentum $p_{\mu}$ we can determine the range of
$q_0$ for which the step functions in Eq.~(\ref{cross_sec4}) do not
vanish. This procedure determines the bounds on $\cos \theta_q$
at fixed beam energy $E_e$, from which bounds on $\phi_q$ follow. The
numerical evaluation of the multiple integral is then performed using
standard quadrature methods.  At the low $Q^2$ values involved here
the $Q^2$-dependence of $G_E$ and $G_M$ induces a correction of
5--10\% in the cross section, as compared to using their $q^2=0$
values. We have taken this into account in the numerical results
presented in Sec.~\ref{sec-EMcrosssection}, using a standard dipole
parameterization obtained from studies of $e^-p$ scattering:

\beq G_E(Q^2)=\frac{G_M(Q^2)}{1+\kappa}=\frac{1}{\left(1+Q^2/0.71 {\rm
GeV}^2\right)^2}\,\quad .  
\eeq
 
For the range of $Q^2$ relevant to the process considered here, this
parameterized form is accurate to better than 1\%.
\end{appendix}

\section*{Acknowledgments}

We acknowledge  valuable conversations with Ken Hicks, whose
ideas regarding EMU stimulated this research. We also thank Vincenzo
Cirigliano for useful discussions on beyond-the-Standard-Model
operators.  This work was supported by the Department of Energy under
grant DE-FG02-93ER40756, and by the Ohio University Office of Research.

\end{document}